# The Role of Arteriovenous Graft Curvature in Haemodynamics: an Image-Based Approach

Guanqi Wang[1], Thomas Abadie[1], Ivan Wall[2], Patricia Pérez Esteban[1,*]

[1] *School of Chemical Engineering, University of Birmingham*
[2] *Institute of Immunology and Immunotherapy, University of Birmingham*
\* Corresponding author at: Translational Medicine, Heritage Building, Mindelsohn Way, Birmingham, B15 2TH, United Kingdom.
E-mail address: p.perezesteban@bham.ac.uk.

**Abstract**

Vascular access, such as arteriovenous grafts, is crucial for patients undergoing haemodialysis as part of kidney replacement therapy. One of the primary causes of arteriovenous graft failure and loss of patency is disordered blood flow, as the vein is exposed to the arterial environment with high flow rates and shear stress. We hypothesize that secondary flow downstream of the vein-graft anastomosis plays a critical role in generating low shear regions, thereby promoting neointima hyperplasia. The secondary flow highlighted here also promotes high oscillatory shear index regions downstream of the vein-graft anastomosis, further contributing to graft failure. To prolong the overall graft survival and patency, we aim to develop a strategy to optimise graft configurations with reduced levels of disturbed haemodynamics. We developed an image-based approach to build three-dimensional geometries for subsequent computational fluid dynamics (CFD) numerical simulations. This simple, yet accurate, method allowed us to improve the accuracy of geometries, thus facilitating comparisons between different vein-graft anastomotic angles.

Our results reveal that overall graft curvature (looped *vs.* straight) plays a dominant role in characterising the failure metrics. Looped grafts, particularly at moderate vein-graft anastomotic angles (30°-45°), exhibited the most favourable metrics, including reduced values of low wall shear stress, high wall shear stress, and high oscillatory shear index. These findings provide critical insights to inform medical professionals about graft areas that are subject to high shear stresses due to the oscillating nature of blood flow as well as the graft geometric configuration when performing surgery. The model developed in this work offers a framework enabling personalised vascular access strategies tailored to individual patient needs.

***Keywords:*** *Computational fluid dynamics; image analysis; OpenFOAM; haemodynamics; arterio-venous graft.*

## 1. Introduction

Haemodialysis is a crucial lifeline for patients with end-stage renal disease (ESRD) who have lost the normal function of one or both their kidneys. Surgically created arterio-venous fistulas (AVF) and arteriovenous grafts (AVG) are recommended options for vascular access worldwide. These surgical interventions allow to connect patients to an external dialysis machine that filters the build-up of fatal toxins in the body. AVGs are often implanted based on the patient's vascular condition and the surgeon's preference, with one end stitched to the artery and the other end to the vein. Unfortunately, AVG dysfunction is common, primarily at the vein-graft anastomosis [1-4], due to complications such as thrombosis (55 %), anastomotic obstruction/stenosis (22 %) and infection (10 %) [5]. Indications of graft dysfunction may include high venous resistance, poor arterial inflow, and increased dialysis time [6, 7]. As of 2023, the primary patency rate – the period a graft remains patent and unobstructed after initial placement – typically remains below 70% at 1 year [8, 9].

Although therapies are available to maintain the graft patency [5, 10, 11], it is uncertain how to prevent these complications to reduce the risk of graft failure. The prevailing theory suggests that the aberrant fluid-wall shear stress (WSS) plays a vital role in promoting venous neointimal hyperplastic stenoses [12-14]. *In vivo* animal studies also indicate a close correlation between aberrant haemodynamics and the development of intimal hyperplastic thickening [15-19]. In the arteriovenous access, shunting of arterial blood flow straight into the vein dramatically disrupts the haemodynamics in the vein, causing increased turbulence and even recirculation eddies, which can ultimately stimulate intima-media thickening and increase the risk of thrombosis secondary to venous hyperplasia [14].

Given the evidence from early long-term animal studies by Fillinger *et al* [15-17], graft geometry is a critical factor influencing local haemodynamics and eventually impacting the AVG outcomes [15-17]. Computational fluid dynamics (CFD) has emerged as a powerful tool to clarify the detailed flow characteristics in AVG models that are beyond the capabilities of experimental measurements. For example, idealised AVG models have been used to understand the essential features of real flow before tackling patient-specific simulations [20-23]. More research aims to optimise AVG configurations, such as adjusting venous anastomotic angles to reduce the turbulence strength and preserve graft function [24-30]. The correlations between the failure areas have rarely been explored, however. One key phenomenon, the Dean vortex, occurs when fluids flow through curved vessels, such as at graft-to-vein bend, generating secondary swirling flows that can contribute to endothelial stress and vascular wall remodelling [31]. To our knowledge, no studies in arteriovenous haemodynamics have explicitly examined the impact of curved bends or the potential correlations between failure metrics and such geometric features.

While patient-specific models may be preferred since they reduce biases when compared to simplified models [32-36], they likely involve multiple independent variables that complicate cross-study comparisons. A recent study by the US Food and Drug Administration highlighted significant variability in CFD results for idealised medical devices [37]. As a consequence, despite advancements, the patency rates of AVG implants have not significantly improved [8, 9], and clinical adoption of novel graft designs remains limited [38]. Best practices in CFD, such as improving methodologies for configuring geometry, have been suggested to enhance regulatory consistency [39].

With the aim to generate tangible outputs that can inform medical professionals of the most appropriate graft configurations, both straight and looped AVG types were modelled in this work. Specific flow features, such as recirculation zones, were examined to determine whether their behaviour changed significantly between configurations. Fibre-based physical models were crafted to mimic *in vivo* or *in vitro* experimental scenarios. A MATLAB-based custom programme was used to digitise these physical models for subsequent numerical simulations. This image-based approach optimised the computer-aided design process and lays a foundation for developing patient-specific models based solely on real medical scans to support clinical decision-making.

## 2. Methodology
### 2.1. Realistic geometry and mesh

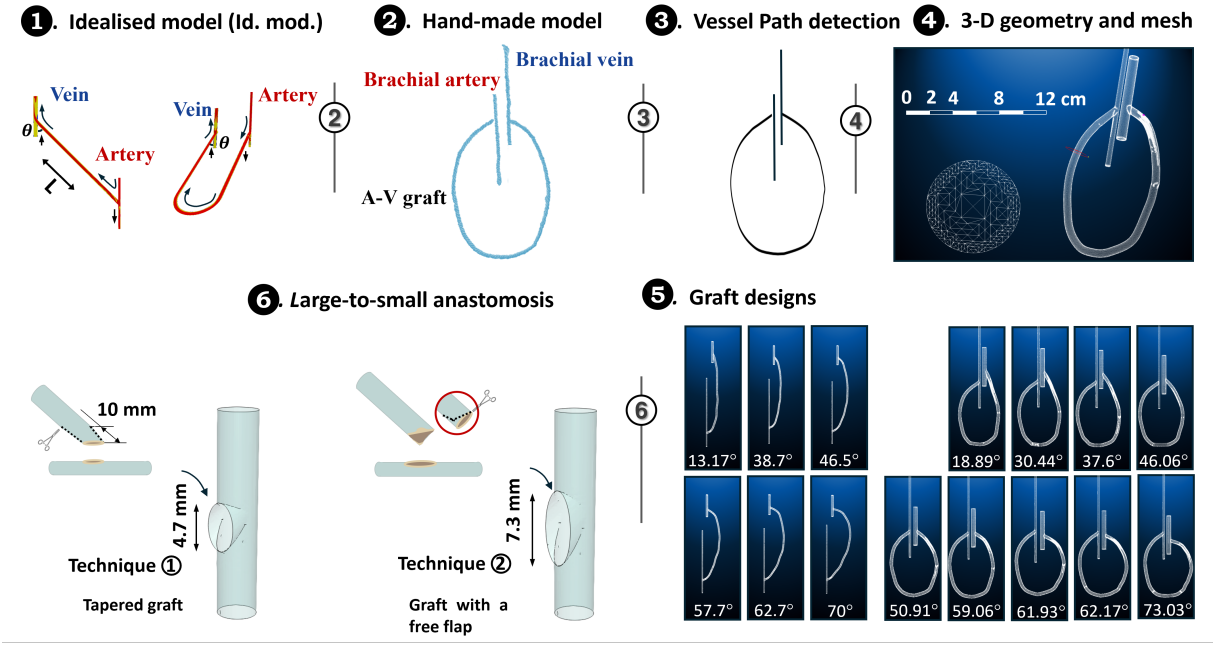

Figure 1. Methodology for constructing a realistic arteriovenous graft model. Panel 1 presents the idealised models. Panel 2 illustrates a hand-made model of the arteriovenous graft, created using fibres, with one end stitched to the brachial artery and the other to the brachial vein. Panel 3 displays the centrelines of the hand-made models, detected by a custom MATALB programme. This provided the 'data path' to generate 3D geometries in OpenFOAM as shown in Panel 4. The utility *blockMesh* can build up vessel pipes, and *surfaceMeshTriangulate* can export geometries in STL format. In all configurations in Panel 5, the fibres representing the graft had a fixed length of 310 mm. Panel 6 shows two techniques for creating the artery-to-graft anastomosis: One method produced a conical tapered segment at the graft end to the artery, extending approximately 10 mm in length, while the other, simulating a surgeon's technique [40], generated an enlarged intersectional area using a free flap.

Adjustable fibres were used to build the network of arteriovenous graft blood vessels (see Panel 2 in Figure 1). This approach closely mimicked *in vivo* surgical operations or *in vitro* experimental scenarios. When the vein-graft anastomosis angle was carefully varied, the artery-graft anastomosis angle remained unchanged, whereas the graft curvature on the venous side adjusted naturally as a consequence (see Panel 5 in Figure 1). This method offers a valuable complement to purely computer-aided design (CAD) approaches. The use of the same physical fibre allowed for easy control of a fixed graft length across all configurations, which is often difficult to achieve in CAD when curvature changes affect overall dimensions [23, 27, 28].

Subsequently, images of different configurations were captured using a smartphone camera, and a custom MATLAB programme was then used for image processing to detect the centrelines of the blood vessels (see Panel 3 in Figure 1). The processed dataset was then implemented in OpenFOAM which subsequently provided the 'path' to generate the 3D curved vessels as shown in Panel 4. For each instance, the artery had an inner diameter of 4 mm (denoted as $D_a$ in Figure 2), and the vein's inner diameter was 8 mm ($D_v$ in Figure 2). The 6 mm graft ($D_g$ in Figure 2) was either tapered down to 4 mm or modified into a free large flap at the arterial end (see Panel 6 in Figure 1), following surgeons' guidance for complex large-to-small anastomosis [40]. Using the latter technique, the anastomosis major axis for looped graft designs reached up to 7.28 mm. In contrast, the tapered graft design leads to reduced flow rates in access graft due to the narrowed intersectional surface area between the artery and the graft [41, 42]. Moreover, scar formation around the connection would further block access graft and may cause clotting, making it less favourable from a clinical perspective.

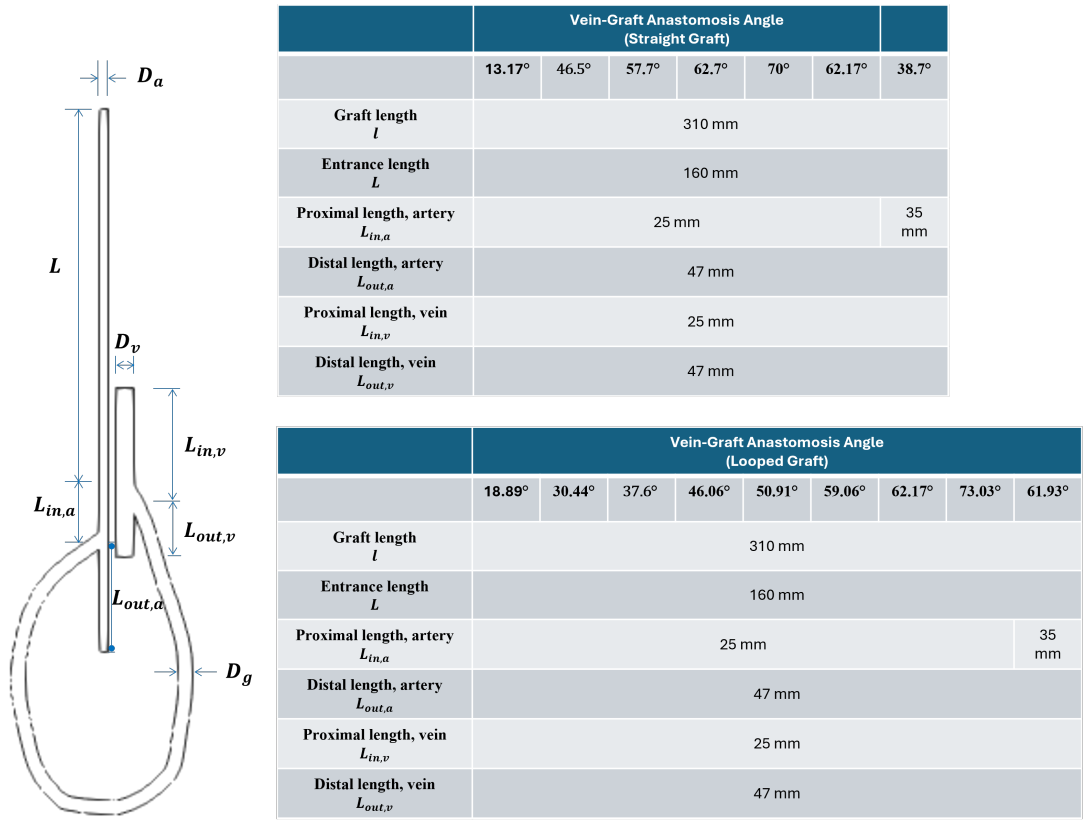

Figure 2. Key parameters of the straight and looped physical models

Vein-graft anastomosis angles were chosen as an essential strategy for graft revision: 13.17°, 38.7°, 46.5°, 57.5°, 62.7°, and 70° for the straight type, and 18.89°, 30.44°, 37.6°, 46.06°, 50.91°, 59.06°, 61.93°, 62.17°, and 73.03° for the looped graft. Key parameters used to define the physical configurations are illustrated in Figure 2 and inset tables. In all simulations, both arterial and venous lengths were standardised to 72 mm, with 47 mm distal and 25 mm proximal to the anastomosis centre. In a few cases, such as the straight graft type with a 38.7° and the looped graft type with a 61.93°, the arterial distal segment was extended slightly, up to 35 mm, to allow the flow to readjust to a stabilised condition, where a uniform outlet boundary condition could reasonably be applied.

Idealised models (Id. mod.) were built up separately for benchmark studies (Panel 1 in Figure 1). These geometries, with lengths of 200 mm, 300 mm, and 400 mm, consisted of simple straight cylindrical segments and regular half-circles. The angle ($\theta$) varied between 10 ° and 60° in 15-degree intervals. Two idealised looped grafts with a length of 400 mm were additionally constructed with $\theta$ set at 10 ° and 45°, respectively. For simplification, the artery-graft anastomotic angles changed as the vein's end was adjusted.

The geometry space was discretised using OpenFOAM (version 10) [43] with an unstructured grid for the simulations. The inset Figure 1, Panel 4, displays a sliced view of the mesh in the cross section of the graft downstream of the graft-artery anastomosis. The OpenFOAM utility *snappyHexMesh* was used to refine the mesh by subdividing a hexahedra (hex) cell into split-hexahedra (split-hex) cells, with particular emphasis on the vicinity of the vessel wall to capture the rapidly changing velocity field in the boundary layer due to no-slip conditions. This was accomplished by capturing regions with significant curvatures to preserve mesh distortion. With the purpose to achieve a high-quality mesh with a good compromise of accuracy and resource consumption, different levels of space discretization were subsequently assessed. Based on the gradually stabilised flow patterns observed in Figure A.2. (see Supplementary Material), mesh level 3 was adopted for all simulation studies. This level contained a total of 573,272 hexagonal cells for the entire graft model. Note, this referenced a looped graft model with a graft-vein anastomosis angle of 59.06°.

The transient nature of blood flow throughout the cardiac cycle was another factor that impacted convergence. Figure A.2. (b) shows that the first heart beating cycle was sensitive to initial conditions but the second, third and fourth beating cycles were identical, which implies that at least two beating cycles are necessary for convergence.

## 2.2. Boundary Conditions

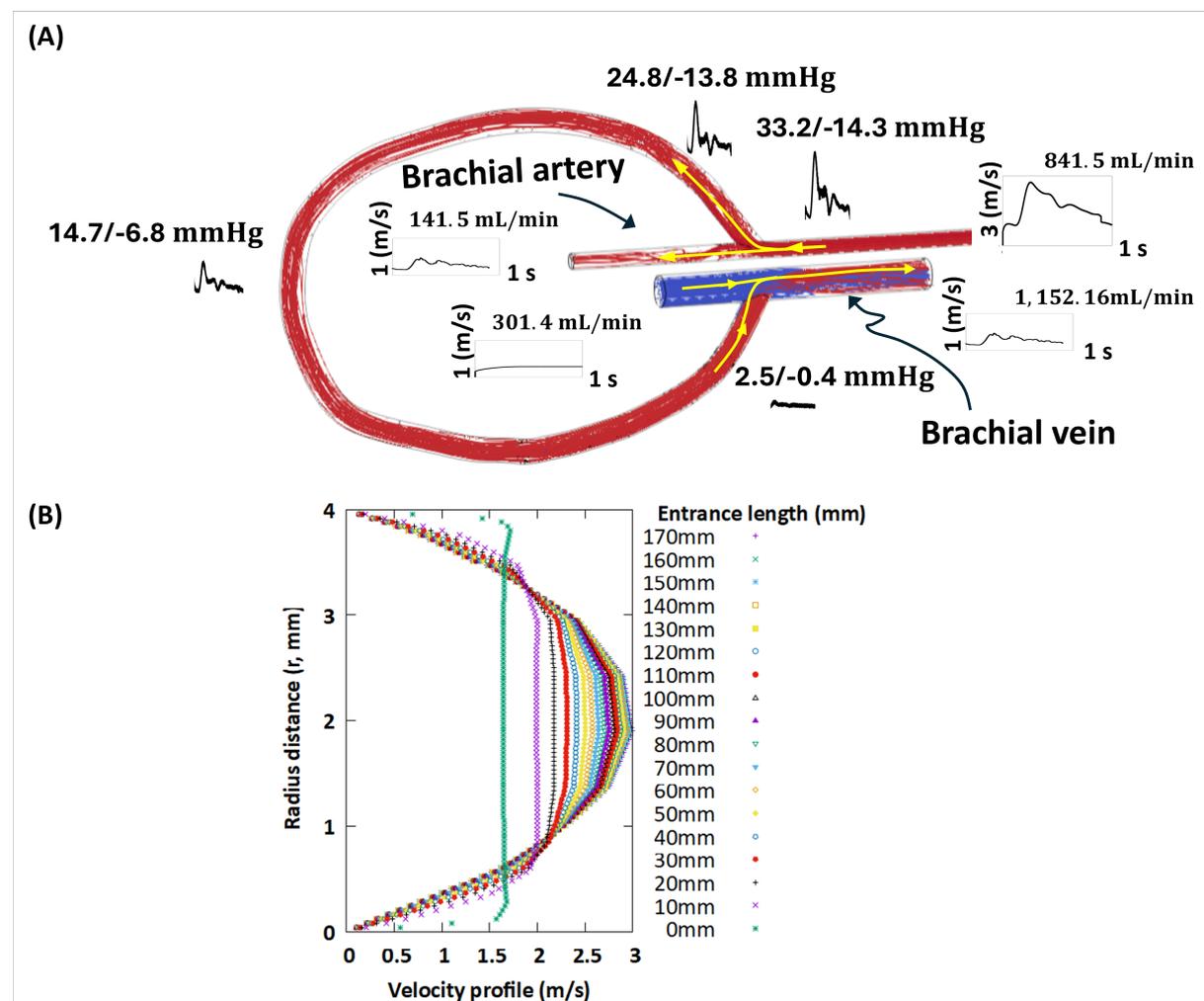

Figure 3. Typical pulsatile flow waveforms as boundary conditions. In panel (A), pressure waveforms are measured results following the computational fluid dynamics analysis, so is the blood flow rate curve at the venous outlet patch. Streamlines are used to visualise the flow behavior throughout the entire model domain, produced by ParaView 5.12.0. Panel (B) displays the developing blood flow profile as the flow enters the domain from the arterial inlet patch.

Figure 3(A) depicts the simulated haemodialysis scenario, where blood flows from the artery, through the graft, and into the vein. At the arterial inlet to the domain, a uniform wave flow was applied (*codedFixedValue* boundary condition in OpenFOAM® 10). The waveform used here was derived from measurements taken in the brachial artery of 19 healthy individuals from the literature [44]. Curve fitting in MATLAB® generated a complete Fourier series with eight harmonics to define the waveform, yielding an average volumetric flow rate of 841.5 mL/min entering the domain. Note, an extended inlet length was required for flow to be fully developed before it entered the domain – the so called 'entrance length'. The entrance length study in Figure 3(B) indicates that an entrance length of approximately 160 mm was required for the 4 mm-diameter artery-like tube. The angular frequency of the first harmonic of the Fourier series of the oscillatory flow is 3.326, corresponding to a Womersley number of $\alpha \approx 1.68$. At this Womersley number, the fully developed profile in Figure 3(B) is quasi-parabolic.

At the venous inlet patch, a constant flow rate was applied given the pulse waves generated by the heart are progressively dampened in the arteries and capillaries [45]. A parabolic profile was specified, to prevent further extending the model geometry, the number of mesh cells, and the computational requirements. The streamwise velocity with respect to radial distance from the vein center was defined as:

$$v = 2v_{(av)} \left[1 - \left(\frac{r}{R}\right)^2\right] \tag{1}$$

where, $R$ represents the vein radius, $r$ denotes the radial distance, and the average velocity is given by $v_{(av)} = 0.15$ m/s.

Outlet boundary conditions are recognised as a significant challenge for vascular flow simulation [24, 32]. Albeit limited data resource downstream of the vascular access, the flow rates in access grafts were obtained from prior studies [46]. For this reason, a volumetric flow rate was defined at the arterial outlet patch by substracting the flow rates in access graft from the entire inlet flow (termed *flowRateOutletVelocity*). Assuming the flow waveform at the arterial outlet patch has the same harmonics as the upstream flow from the inlet patch, the volumetric flow rate was correspondingly defined as:

$$q(t) = f \times \left(a_0 + \sum_{n=1}^{8} a_n \cos(nt) + \sum_{n=1}^{8} b_n \sin(nt)\right) \tag{2}$$

where, $a_0$, $a_n$ and $b_n$ are curve fit parameters inherited from the arterial inlet function, $t$ is the time, and $f$ denotes the volume fraction from the upstream flow. The straight graft exhibited lower flow rates (517.5 mL/min) compared to a looped graft (700 mL/min), regardless of the vein position [46]. Therefore, the volume fraction was set to $f = 0.168$ for the looped graft studies and 0.385 for the straight graft studies.

### 2.3 Simulation setup
Blood flow simulations were treated as a transient flow of a viscous, incompressible fluid. Given the vessel size was larger than blood cells ($\sim 10^{-3}$ cm) and shear rate was sufficiently large to neglect blood rheology, the fluid was assumed to be Newtonian, and therefore with a constant dynamic viscosity of 0.005 Pa•s and a constant density of 1060 kg/m³ [47]. The finite volume pisoFoam solver in OpenFOAM 10 was subsequently selected to solve the Navier-Stokes equations [43], which is a transient solver including time derivatives to account for the changes in velocity and pressure over time. Gravity effects were neglected because: (1) flow in the arterio-venous grafts is primarily pressure-driven, and (2) the A-V grafts and associated vessels are small in size, making hydrostatic pressure negligible compared to viscous and inertial effects. Aligned with the boudnary curves defined in Figure 3(A), the flow velocity at the arterial inlet, as averaged over the cross section, was $U_{(av)} = 1.12$ m/s. This yields a Reynolds number of

$$Re = \frac{\rho D U_{(av)}}{\mu} = 949.76 \tag{3}$$

The graft flow rate was lower due to a smaller volume flow rate and a larger cross-sectional area. The venous flow rate was considerably low at 0.15 m/s, which indicates the flow was strictly under laminar condition.

The discretized Navier-Stokes equations were iteratively solved, with the velocity and pressure fields being considered to achieve convergence when the absolute residuals fell below a tolerance criteria of $1 \times 10^{-5}$ and $1 \times 10^{-6}$ respectively. As an example, for the looped graft model with vein-graft anastomotic angles of 59.06°, the minimum cell volume in the mesh was $1.29516 \times 10^{-13}$ m³, and the minimum face area was $3.02117 \times 10^{-10}$ m². This yielded a minimum cell length in the flow direction ($dx$) of maximum $4.3 \times 10^{-4}$ m. Blood flow would pass through this cell over a characteristic time of $dx/(2U_{(av)}) = 1.9 \times 10^{-4}$ s. To ensure the governing equations was resolved with the required accuracy, a time step of $1 \times 10^{-5}$ s was used for consistent convergence of solutions. Futher details on the numerical methods and solver settings is provided in Appendix A.1 for readers who wish to explore the technical details.

**2.4. Key metrics for hemodynamic insight**

Equations (4)-(6) define the key metrics to quantify the wall shear stresses that deviate from the normal physiological ranges of 0.1 to 7 Pa in the vein segment [48]. Currently, there is no consensus on a specific threshold for aberrant WSS, as previous *in silico* studies have utilized low WSS values varying from 0 to 3 Pa [49-51]. In this study, the threshold values were selected according to average measurements in the venous (0.1 – 0.6 Pa) and arterial systems (1-7 Pa), respectively [48]. Denoting $\lambda$ as the measured low WSS and $\dot{\lambda}$ as the low WSS threshold (0.1 Pa), which indicates a risk of neointimal hyperplasia, the Low WSS Risk Gradient Metric, $\Lambda_{low}$, was defined to quantify the extent and severity of low WSS. This metric was calculated as the product of the area with low WSS, $\alpha$, and the degree of shortfall from the threshold $\dot{\lambda}$, normalised by the total area of the vein wall, $A_0$:

$$\Lambda_{low} = \left(\frac{\dot{\lambda}}{\lambda} \cdot \alpha\right)/A_0 \tag{4}$$

Here, $\dot{\lambda} = 0.1$ Pa is the threshold value, and $A_0 = 0.00181$ m² is the total area of the vein wall.

Similarly, the High WSS Risk Gradient Metric, $\Gamma_{high}$, was defined to assess the risk associated with high WSS. Let $\gamma$ represent the measured high WSS value, and $\dot{\gamma}$ the upper threshold for high WSS (7 Pa). This metric was calculated as the product of the area exposed to high WSS, $\beta$, and the extent by which the measured $\gamma$ exceeded the threshold $\dot{\gamma}$, standardized by the area of the entire vein wall, $A_0$.

$$\Gamma_{high} = \left(\frac{\gamma}{\dot{\gamma}} \cdot \beta\right)/A_0 \tag{5}$$

where, $\dot{\gamma} = 7$ Pa is the threshold value. Note that $\Lambda_{low}$ and $\Gamma_{high}$ reflect measurements taken at each time step of the simulation.

Equation (6) defines the Oscillatory Shear Index (OSI) as a measure of the directional changes in WSS in response to the pulsatile nature of the blood flow. The OSI was used to compare the magnitude of the mean WSS vector and the mean magnitude of the WSS over a cardiac cycle. It quantified the extent to which WSS oscillated back and forth over a cardiac cycle, providing insights into the disturbed or potentially pathological flow within the vein segment.

$$OSI = \frac{1}{2}\left(1 - \frac{\left\|\int_0^T \vec{\tau}(t)\,dt\right\|}{\int_0^T \|\vec{\tau}(t)\|\,dt}\right) \quad (6)$$

where, $\vec{\tau}(t)$ is the WSS vector at time $t$ within the vein segment. To ensure the average was calculated within the same cardiac cycle, the field average was set up over time steps with a periodic restart every 0.9 sec, correponding to the duration of the cardiac cycle. OSI values near 0 indicate unidirectional (healthy) flow, while values approaching 0.5 suggest significant oscillations in flow direction, which are typically associated with disturbed or turbulent flow conditions that promote thrombosis. A threshold value of OSI > 0.25 was used to identify regions subjected to highly oscillatory WSS [32].

## 3. Results

The wall shear stress values fluctuate in response to the pulsatile nature of blood flow in the brachial artery. The focus here is primarily on conditions in the vein segment, as that is where the stenosis and thrombosis mainly occur [1]. To illustrate the nature of the wall shear stress, Figure 4 shows the shear stress countour maps on the wall patches. A positive association was observed between the wall shear stress values and the blood flow waveforms. The vessel wall encountered high percentage of pathological patches (in red color) when the blood flow rate is high. These problematic regions, however, greatly fade away at the trough of low blood flow rate. In more detail, among the different configurations small vein-graft anastomosis led to reduced pathological regions under the effects of high wall shear stresses, *i.e.* vein-graft anastomosis at 13.2° in Figure 4. To further quantify this correlation, the low-risk regions was filtered in post-processing operations and applied equations (4)&(5) over nine successive time points in a complete cardiac cycle. In Figures 5 (a-*i*),(b-*i*)&(c-*i*), the initial decrease of the Low Risk Gradient Metrics arose from the acceleration of blood flow in the early phase of a circulatory loop. Since blood flow fell in the second phase, there was a remarkable increase of the Low Risk Gradient Metrics at the final time point, associated with the lowest blood flow rate measurement. It is worth noting that the risk regions varied over time, and moreover, at some time points no risk patches were observed under low wall shear stresses. The time-varying risk metrics differed by the order of magnitude of 0 to 2 on the dependance of the graft geometry and vein-graft anastomosis angles. In Figures 5(a-*ii*),(b-*ii*)&(c-*ii*) , the peak of High Risk Gradient WSS occurs at 2.1 s in association with the highest blood flow measurement among the 9 time points.

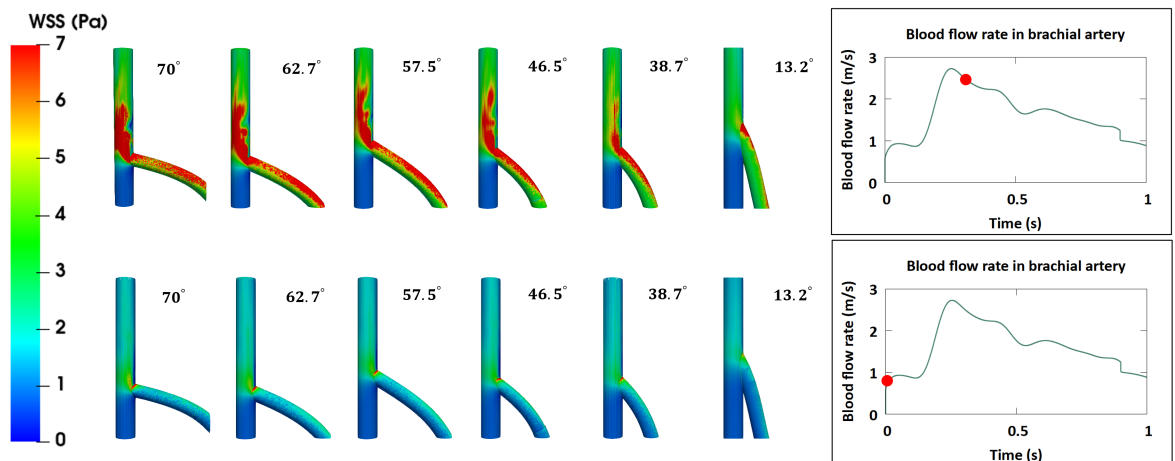

Figure 4. Wall shear stress countour maps classified by the fluctuating blood flow rate. Maps were color-coded according to the magnitude of the full three-dimensional wall shear stress, as shown by the color bar. To accentuate the risk regions impacted by aberrant wall shear stress, the maximum values on the color bar was chosen based on the threshold value 7 Pa, not on the entire data range. The top panel shows affected patches by high wall shear stress values under high blood flow rate.

Conversely, the bottom panel presents considerably low wall shear stress under low blood flow rate. Geometries here were clipped out of the straight graft series to reflect the vein-graft anstomosis.

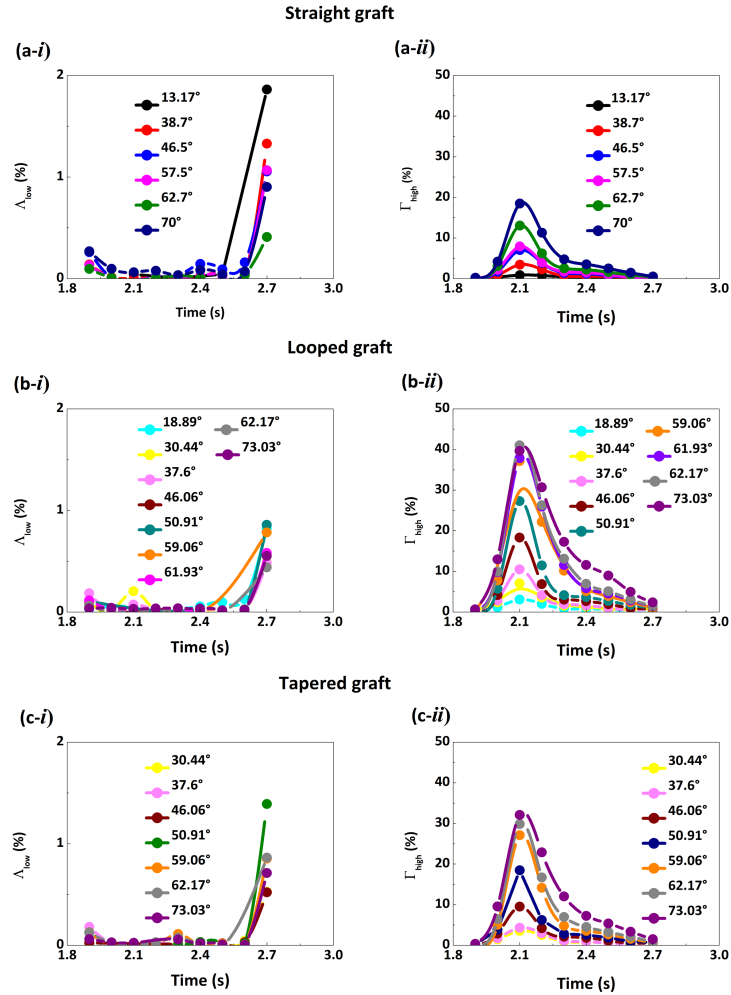

Figure 5. Display of Low (first collumn) and High (second collumn) Risk Gradient WSS metrics over a complete cardiac cycle. Data sampling was conducted over nine time points with a time step of 0.1 s.

Figure 6 depicts the mechanics of flow disturbances at the vein-graft anastomosis over the 9 time points. As known, the magnitudes of Low and High Risk Gradient Metrics would change in response to the time-varying blood flow rate (see panel A). In Figure 6B, one can see a swirl downstream of the blood flow at the vein-graft anastomosis where the two blood streams met. This is known as a Dean vortex, which happens when the fluid flows through a curved pipe such as an elbow [31]. Due to the momentum of the fluid, the blood stream from the graft flew towards the outer wall of the anastomosis bend when joining the venous stream. In addition to a general migration of the velocity profile in the graft inflow, the different magnitude of the venous inflow added more complexities to the velocity gradient. This generated a pressure gradient in the inner radial direction (see panel C) that gave rise to a secondary motion superposed on the primary flow in the distal vein direction. A clear Dean vortex was visible at slow arterial blood flow, *i.e.* $t_1$, $t_2$, and $t_9$, yielding a high risk gradient $\Lambda_{low}$. Whereas the secondary flow became unstable at high arterial blood flow, *i.e.* $t_3$-$t_6$, which accounts for the corresponding high risk graident $\Gamma_{high}$. This instability can also be traced in $y$ space: the flow became considerably disordered at high arterial blood flow rate (see red streamlines in panel B). In the outer radial direction, the pressure gradient was not significant (see panel C), and therefore the flow here was more unidirectional as displayed in Figure 6B.

The inset panels in Figure 6B traced blood streams originating from the graft. The curvature of the bend was determined using three featured points along the flow path: the entry point $P_1$, midpoint $P_3$ and the exit point $P_2$. At time point $t_1 = 1.9$ s, the upstream blood flow velocity at the vein-graft junction was approximately ~0.156 m/s. The curvature of the bend was estimated to be 45.6, yielding a Dean number of $De \sim 99$. At time point $t_3 = 2.1$ s, the blood flow rate increased to 0.45 m/s, and the curvature decreased to about 25 due to the changing flow path. This increased the Dean number significantly to $De \sim 210$, which resulted in unstable secondary flows and corresponding high risk metrics $\Gamma_{high}$.

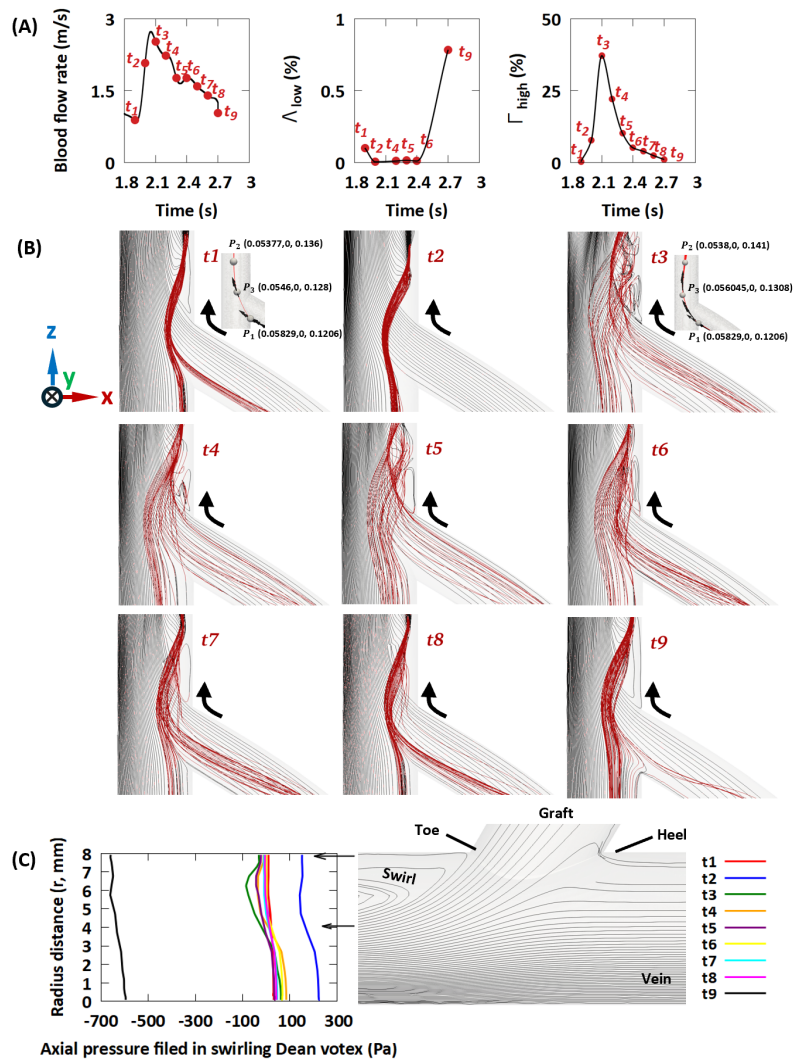

Figure 6. The mechanics of flow disturbances at the vein-graft connection. Panel (B) is a showcase of the flow streamline trace in the $x$-$z$ plane, with the inner radius in the positive $x$ diretion and the outer radius in the negative $x$ direction. Due to the flow disturbances at the junction where the two streams met, the trace demonstrated some inflow arising from $y$ space. By clipping the geometry at $y = 0 \,(\pm 1)$ mm, the black streamlines characterised the flow morphology in $x$-$z$ space. The red streamlines represented the other portion that was clipped out. Panel (C) shows a radial pressure gradient close to the downstream of the anastomosis. This gave rise to a secondary flow, known as Dean vortices. $t_1$-$t_9$ correspond to the nine time points from 1.9 sec to 2.7 sec. Example for a loop graft design with a graft-to-vein anastomosis angle of 59.06°.

The secondary flow also means slow moving blood near the inner side of the anastomosis, that is, the blood flow rate near the inner wall was smaller than that in the outer side. Herein, low wall shear stresses were observed in the preceding and receding regions of the Dean swirl downstream of the anastomosis (see Figure 7). This is in line with the clinical discoveries: the neoinitimal hyperplasia often occurs downstream of the graft-vessel anastomosis. When the arterial blood flow dropped off to the lowest point, these problematic patches spread to the bottom wall, and surprisingly, over an increased distal distance. Meanwhile, Figure 6C shows a sudden reduction in pressure at time point $t_9$.

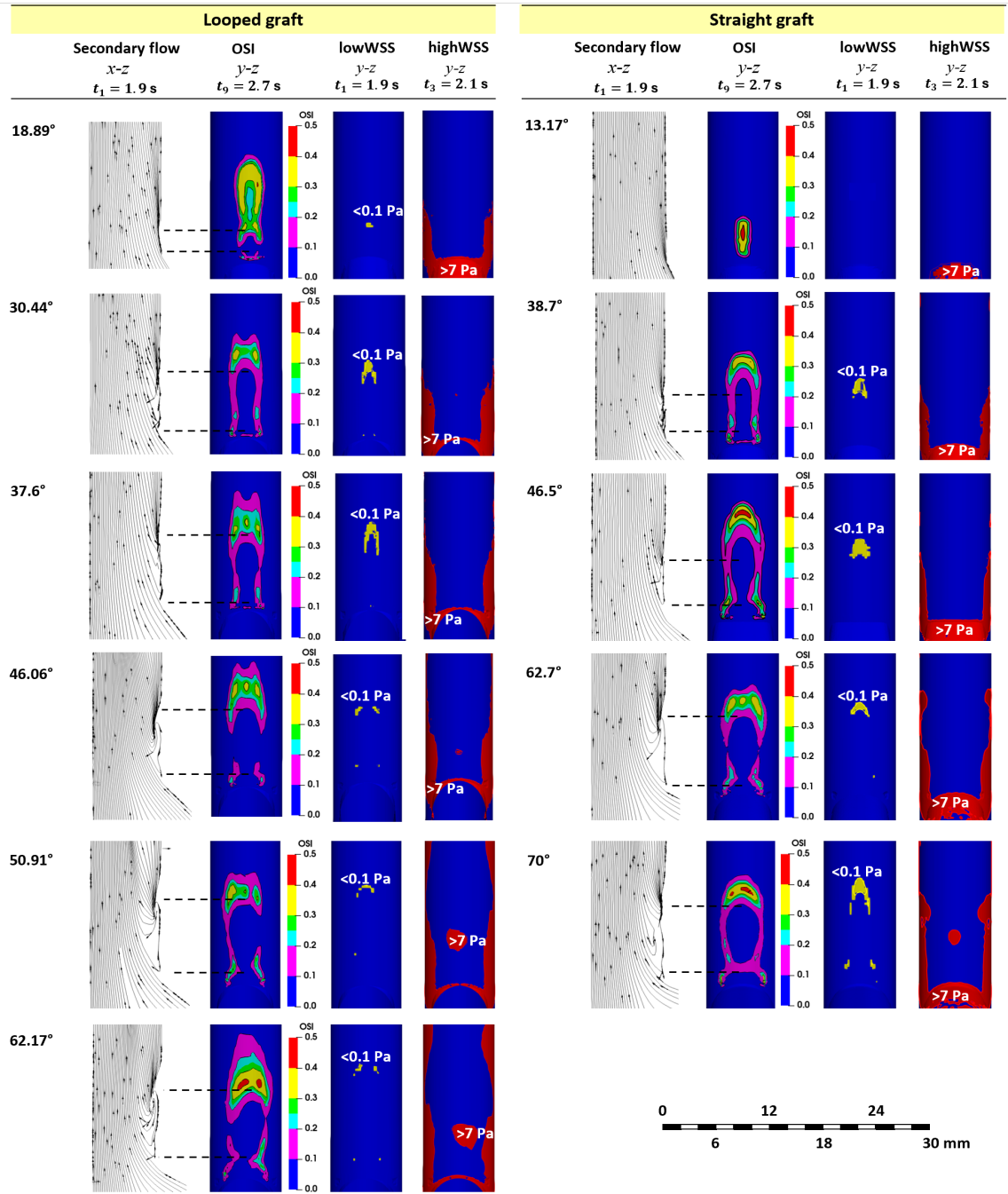

Figure 7. Problematic flow regions in relation to the Dean secondary flow.

Figure 7 demonstrates that the pathological flow fields were mostly distributed downstream of the vein-graft anastomosis over a limited distance within 50 mm. Notably, high oscillatory flow ($OSI > 0.25$) [32] was observed in the inner bend channel, in close proximity to the Dean vortex, and extended further into the downstream region. The oscillations were likely induced by the complex interaction between the secondary swirl flow and the primary flow. However, the flow immediately surrounding the swirl core was relatively undisturbed, expecially in the axial direction of the vein vessel. While more disturbances occurred in the radial direction, they were less severe ($OSI < 0.2$). This indicates a region of relatively stable flow winthin the swirl, where the velocity gradients and secondary motion did not propagate significantly outwards, despite the presence of secondary instablilities at high arterial blood flow rate as presented in Figure 6B.

Figure 7 further illustrates that the Dean vortex enlarged at larger vein-graft anastomotic angles for both looped and straight graft types due to a reduced radius of curvature. In general, looped graft generated larger swirls due to higher velocity scales and greater graft inflow access. When the vein-graft anastomotic angle was below 20°, straight grafts exhibited predominantly unidirectional flow throughout the entire cardiac cycle. Looped grafts, however, continued to develop secondary flows but at higher arterial blood flow rates due to their greater graft inflow access.

In Figure 7, regions of high wall shear stress were primarily located preceding the Dean vortex and in the side wall patches of the bend. At larger vein-graft anastomotic angles, high wall shear stress values were also observed at the center of the Dean vortex, implying the enlarged Dean vortex was associated with an increased mean flow rate.

In Figure 8, the risk metrics were further analysed in relation to the Dean number $De$. Figures 8 (a)&(b) illustrate this correlation study at the classical time points $t_1$ and $t_3$. In both Figures 8 (a)&(b), regions of high WSS ($\Gamma_{high}$) increased with the Dean number, with a steeper curve slope for $De > 100$. Figures 8 (c)&(d) present analogous results for the straight type studies, with slight variations in the correponding curve slopes.

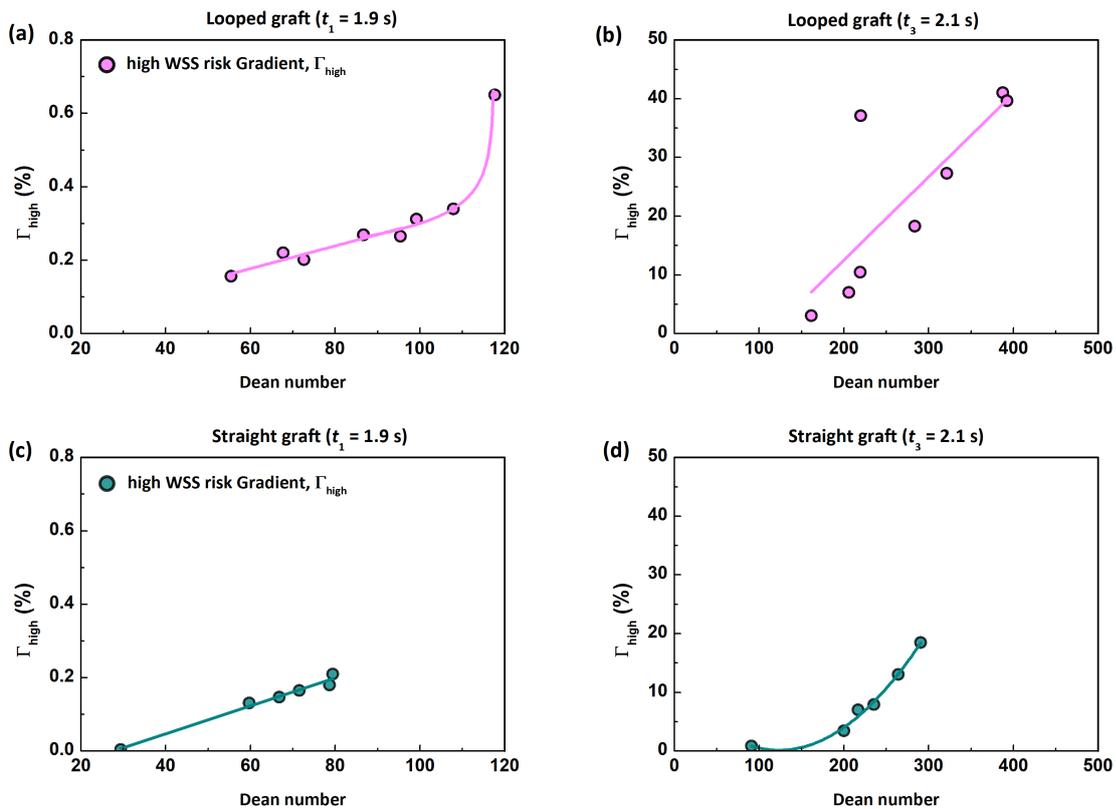

Figure 8. A correlation study between the risk metrics and the Dean number.

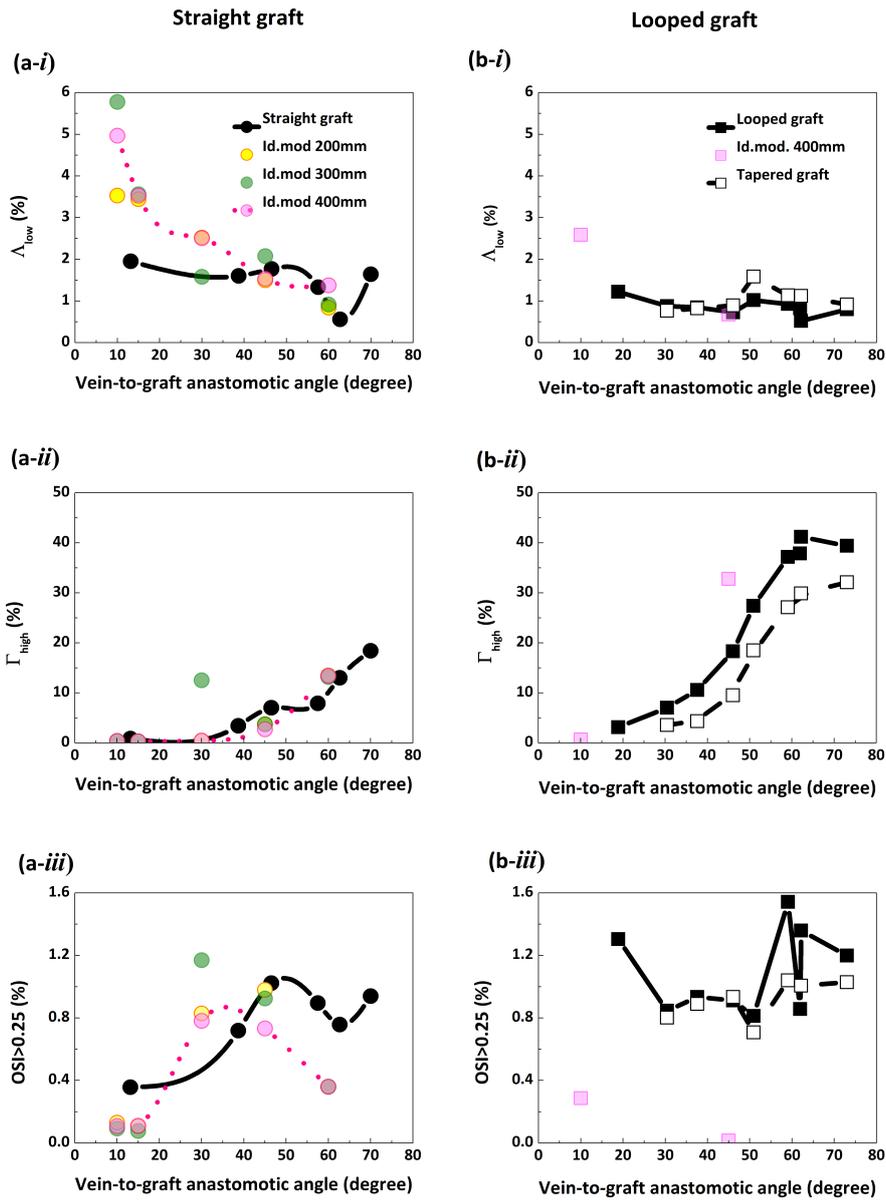

Figure 9. Summary of risk metrics in straight and looped graft configurations. Short dashed lines in panels (a), (c) & (e) are curve fitted results based solely on the data from 400mm-graft.

Figure 9 summarizes the risk assessement across the three graft types. To account for the accummulation process when fibrin builded up at slow moving blood field, an integral of $\Lambda_{low}$ over all time points in the entire aortic cycle was summarised here. Figures 9 (a-*i*)&(b-*i*) show that straight grafts experienced more high risk regions impacted by low WSS. This is because straight grafts normally have reduced flow rates due to high flow friction caused by the reversal of stream split at the arterial end [46]. The lower blood flow rate at vein-graft anastomosis likely accounts for the increased percentage of slow-moving blood in straight grafts. Albeit slight oscillations at large anastomotic angles, the risk metric curve trends for both configurations generally demonstrate a decreasing fashion. In Figures 9 (a-*iii*)&(b-*iii*), the oscillatory index numbers were highly comparable for all graft types, particularly at large anastomotic angles (45° to 75°). Below 45° anastomotic angle, looped grafts were detected with more high-risk regions impacted by highly oscillatory flow patterns.

Figures 9 (a-*ii*)&(b-*ii*) show their distinct responses to varying vein-graft anastomotic angles in terms of high risk gradient metrics. Since high wall shear stress causes endothelial cell dysfuncion, the strongest wall shear stress values were used for subsequent configuration comparisons. Generally, both graft types showed a rising trend with larger vein-graft anastomotic angles which is in a good agreement with recent studies by Williams *et al*. [24] and Pahlavanian *et al*. [30]. However, a notable difference in the magnitude is also evident. Looped grafts demonstrated a more pronounced sensitivity to these changes, resulting in significantly higher risk percentages. The risk curve for looped grafts reached a plateau at 60°, whereas the curve for straight grafts continued to rise gradually until it levelled off at 75°.

Interestingly, all the three risk metrics in tapered grafts show a consistent pattern to those in the looped configurations. This is not beyond our expectation given they both were configured in a looped manner. Owing to the reduction in flow rates in tapered grafts [41], high WSS and high OSI fields were reduced whilst low WSS regions were enhanced. This suggests that the overall graft curvature has a major impact on the trend of the risk measurements, but the magnitude of the risk rates is on the dependance of flow rates in access grafts.

Results from the idealised models show a reasonable agreement with the realistic models, albeit with some deviations. Their curve slopes become notably steeper (see Figures 9 (a-*i*)&(b-*i*) for low WSS, 8 (a-*iii*) for OSI, and 9 (b-*ii*) for high WSS). Moreover, the results demonstrate little dependence on the graft length (see Figures 9 (a-*i*)-(a-*iii*)). This is because fluid velocity profiles reach a fully developed regime after a certain distance in a straight pipe, the so-called entrance length. The average velocity in the straight grafts scaled as ∼0.375 m/s, corresponding to a theoretical entrance length of $L = 0.0575 Re_L D_g$~165 mm. Note, the same boundary conditions were applied for the different graft length study given the flow rates in access grafts is mainly determined by the arterial site regardless of the vein position [46].

## 4. Discussion

The primary cause of AVG failure is attributed to the the aberrant heamodynamics arising from the disturbed flow patterns at the vein-graft anastomosis. This study aims to elucidate to what extent these pathological flow fields can be mitigated by adjusting graft curvature and anastomotic configurations. With the purpose to build accurate geometries, an image-based approach was proposed to transform photography images into 3D digitized geometries. Herein, the computer-aided designs for subsequential numberial simulations closely mimicked *in vivo* or *in vitro* experimental designs, allowing elimination of side effects of irrelevant variables that are beyond the scope of the study, *i.e*., artery-to-graft anastomosis angle and graft length. In future, this method enables generation of real patient-specific models based solely on medical scan.

This work examined straight and looped graft configurations since they are both in use in renal centers around the world. The aberrant wall shear stress (WSS) was classified in two categries. On the one hand, high WSS (> 7 Pa) poses a risk of endothelial cell injury and potentially promotes neointimal thrombosis [48]. Risk area analysis in Figure 9 (a-*ii*)&(b-*ii*) showed a rising trend in high WSS regions with larger vein-graft anastomosis angles. This finding is in a good agreement with recent studies by Williams *et al*. [24] and Pahlavanian *et al*. [30], which similarly reported that steeper anastomosis angles led to intensified WSS near the venous outflow. However, a more sensitive response was observed in the looped graft curve, indicating that haemodynamic parameters—such as high WSS regions—are more strongly affected by changes in anastomotic angle in this configuration. On the contrary, straight grafts exhibited a more gradual response. Clinically, this suggests that looped grafts may be more prone to haemodynamic instablitiy under suboptimal angles, potentially increasing the risk of endothelial damage or graft failure. On the other hand, low WSS (< 0.1 Pa) is associated with slow-moving blood, favoring fibrin deposition and leading to stenosis [48]. In this case, risk area analysis in Figure 9 (a-*i*)&(b-*i*) displayed a declined trend with increasing venous anastomotic angles. Interestingly, looped grafts were less sensitive to this trend compared to straight grafts. It is hypothesised that the primary reason for this difference lies in the varying flow rates in access grafts.

The results demonstrate that the haemodynamics in AVGs is significantly influenced by global graft curvature: (i) Straight and looped graft configurations exhibited distinct behaviors when evaluated using risk gradient metrics for low wall shear stress (WSS), high WSS, and osillatory shear index (OSI); (ii) The corresponding curve slopes differed notably between the idealised models and the more realistic hand-made models. This highlights the limitations of using overly simplified geometries to accurately predict flow behavior. Prior studies have attempted to modify local curvatures at the graft-vein end to alter blood flow patterns [57-59]. Building on the findings of this study, it is hypothesized that more attention is needed in the native vessel curvature, as flow disturbances downstream of the problematic vein-graft anastomosis could be further impacted.

A notable feature identified in this work was a Dean vortex downstream of the vein-graft anastomosis. The problematic regions were found closely associated with the Dear vortex regions. Its stability was influenced by not only the pulsatile arterial blood flow rate but also the vein-graft anastomosis angle. The swirl was associated with slow-moving blood near the inner wall which explains the high frequency of low WSS measurements at this location, aligning with clinical observations of adverse outcomes in those regions. A clear Dean vortex occured at the minimum arterial blood flow rate, corresponding to a peak in low WSS failure areas. High oscillatory flow patterns were primarily located adjacent to and downstream of the swirl, whereas high WSS failure areas were concentrated in regions preceding the Dean vortex and along the sidewalls of the bend. Clinically, this suggests that the curved section of the graft may be particularly vulnerable to endothelial damage due to elevated shear stress, potentially contributing to neointimal hyperplasia and long-term graft failure. This highlights the importance ofconsidering graft curvature in surgical planning and design.

The main limitation of this sutdy is the use of a Newtonian fluid model, which does not account for the complex rheological properties of blood. In reality, the relationship between internal forces and velocity gradients becomes non-linear due to the aggregatable and deformable nature of the red blood cells [52]. Altough WSS predictions in small arteries can vary significantly between Newtonian and non-Newtonian models [53], the larger diameter of venous vessels may help mitigate these discrepancies [54]. In the future, a fluid-structure interaction model will enhance physiological realism for additional consideration of the solid mechanics and the interplay between high-pressure blood flow and elastic blood vessels. Further, the use of real medical scan will help to build patient-specific models so as to support clinical decision-making strategies [32].

## 5. Conclusion

Haemodynamics in arteriovenous grafts is significantly influenced by the graft curvature (straight grafts *vs.* looped grafts) (idealised models *vs.* realistic models), vein-graft anastomotic angles and graft modifications at the arterial end (tapered *vs.* non-tapered). The image-based approach is a superior strategy to assist computer-aided designs for complex network of blood vessels. The findings suggest that looped graft configurations with moderate vein-graft anastomotic angles (30°-45°) are most effective in minimizing aberrant flow patterns and therefore have the potential to prolong overall graft patency rates.

**Acknowledgement**

We thank the financial support from the European Union's 9th framework Programme Horizon Europe, under grant agreement NO. 101057673.

**Author contributions**

P.P.E. and I.W. secured the funding acquisition and formulated the project conceptualization. G.W. contributed to the research conceptualization under the supervision of P.P.E. G.W. developed most the methodology and simulation investigations to which T.A. provided technical supervision. G.W. wrote the original draft with inputs, review and editing from T.A. and P.P.E.

**Declaration of competing interest**

The authors declare that they have no known competing financial interests or personal relationships that could have appeared to influence the work reported in this paper.

**Ethics statement**

None.

**Appendix A. Supplementary data**

The following are the Supplementary data to this article:

**References**
[1] R.Y. Kanterman, T.M. Vesely, T.K. Pilgram, B.W. Guy, D.W. Windus, D. Picus, Dialysis access grafts: anatomic location of venous stenosis and results of angioplasty, Radiology 195 (1995) 135-139. https://doi.org/10.1148/radiology.195.1.7892454.
[2] D.E. Finlay, D.G. Longley, M.C. Foshager, J.G. Letourneau, Duplex and color Doppler sonography of hemodialysis arteriovenous fistulas and grafts, Radiographics 13 (1993) 983-989. https://doi.org/10.1148/radiographics.13.5.8210602.
[3] I. Van Tricht, D. De Wachter, J. Tordoir, P. Verdonck, Hemodynamics and complications encountered with arteriovenous fistulas and grafts as vascular access for hemodialysis: a review, Ann. Biomed. Eng. 33 (2005) 1142-1157. https://doi.org/10.1007/s10439-005-5367-X.
[4] T. Lee, N.U. Haq, New developments in our understanding of neointimal hyperplasia, Adv. Chronic Kidney Dis. 22 (2015) 431-437. https://doi.org/10.1053/j.ackd.2015.06.010.
[5] M.E. Cinat, J. Hopkins, S.E. Wilson, A prospective evaluation of PTFE graft patency and surveillance techniques in hemodialysis access, Ann. Vasc. Surg. 13 (1999) 191-198. https://doi.org/10.1007/s100169900241.
[6] K.L. Sullivan, A. Besarab, J. Bonn, M.J. Shapiro, G.A. Gardiner Jr, M.J. Moritz, Hemodynamics of failing dialysis grafts, Radiology 186 (1993) 867-872. https://doi.org/10.1148/radiology.186.3.8430200.
[7] R. Agarwal, J.L. Davis, Monitoring interposition graft venous pressures at higher blood-flow rates improves sensitivity in predicting graft failure, Am. J. Kidney Dis. 34 (1999) 212-217. https://doi.org/10.1016/S0272-6386(99)70345-8.
[8] C.E. Lok, T.S. Huber, A. Orchanian-Cheff, D.K. Rajan, Arteriovenous access for hemodialysis: a review, J. Am. Med. Assoc. 331 (2024) 1307-1317. https://doi.org/10.1001/jama.2024.0535.
[9] R.J. Halbert, G. Nicholson, R.J. Nordyke, A. Pilgrim, L. Niklason, Patency of ePTFE arteriovenous graft placements in hemodialysis patients: systematic literature review and meta-analysis, Kidney360 1 (2020) 1437-1446. https://doi.org/10.34067/kid.0003502020.
[10] P. Bachleda, P. Utikal, M. Kocher, M. Cerna, J. Fialova, L. Kalinova, Arteriovenous graft for hemodialysis, graft venous anastomosis closure-current state of knowledge. Minireview, Biomed Pap Med Fac Univ Palacky Olomouc Czech Repub 159 (2015) 027-030. http://dx.doi.org/10.5507/bp.2014.027.
[11] P.M. Kitrou, K. Katsanos, P. Papadimatos, S. Spiliopoulos, D. Karnabatidis, A survival guide for endovascular declotting in dialysis access: procedures, devices, and a statistical analysis of 3,000 cases, Expert Rev. Med. Devices 15 (2018) 283-291. https://doi.org/10.1080/17434440.2018.1454311.
[12] P.N. Tijburg, M.J. Ijsseldijk, J.J. Sixma, P.G. de Groot, Quantification of fibrin deposition in flowing blood with peroxidase-labeled fibrinogen. High shear rates induce decreased fibrin deposition and appearance of fibrin monomers, Arterioscler. Thromb. Vasc. Biol. 11 (1991) 211-220. https://doi.org/10.1161/01.ATV.11.2.211.
[13] A.H. Gillespie, A. Doctor, Red blood cell contribution to hemostasis, Front. Pediatr. 9 (2021) e629824. https://doi.org/10.3389/fped.2021.629824.
[14] M.K. Fitts, D.B. Pike, K. Anderson, Y.-T. Shiu, Hemodynamic shear stress and endothelial dysfunction in hemodialysis access, Open Urol. Nephrol. J. 7 (2014) e33. https://doi.org/10.2174/1874303x01407010033.
[15] M.F. Fillinger, E.R. Reinitz, R.A. Schwartz, D.E. Resetarits, A.M. Paskanik, C.E. Bredenberg, Beneficial effects of banding on venous intimal-medial hyperplasia in arteriovenous loop grafts, Am. J. Surg. 158 (1989) 87-94. https://doi.org/10.1016/0002-9610(89)90353-X.
[16] M.F. Fillinger, E.R. Reinitz, R.A. Schwartz, D.E. Resetarits, A.M. Paskanik, D. Bruch, C.E. Bredenberg, Graft geometry and venous intimal-medial hyperplasia in arteriovenous loop grafts, J. Vasc. Surg. 11 (1990) 556-566. https://doi.org/10.1016/0741-5214(90)90302-Q.
[17] M.F. Fillinger, D.B. Kerns, D. Bruch, E.R. Reinitz, R.A. Schwartz, Does the end-to-end venous anastomosis offer a functional advantage over the end-to-side venous anastomosis in high-

output arteriovenous grafts?, J. Vasc. Surg. 12 (1990) 676-690. https://doi.org/10.1067/mva.1990.24914.

[18] F. Loth, S.A. Jones, C.K. Zarins, D.P. Giddens, R.F. Nassar, S. Glagov, H.S. Bassiouny, Relative contribution of wall shear stress and injury in experimental intimal thickening at PTFE end-to-side arterial anastomoses, J. Biomech. Eng. 124 (2002) 44-51. https://doi.org/10.1115/1.1428554.

[19] S. Misra, D.A. Woodrum, J. Homburger, S. Elkouri, J.N. Mandrekar, V. Barocas, J.F. Glockner, D.K. Rajan, D. Mukhopadhyay, Assessment of wall shear stress changes in arteries and veins of arteriovenous polytetrafluoroethylene grafts using magnetic resonance imaging, J Vasc Interv Radiol 29 (2006) 624-629. https://doi.org/10.1007/s00270-005-0168-z.

[20] F. Loth, P.F. Fischer, N. Arslan, C.D. Bertram, S.E. Lee, T.J. Royston, W.E. Shaalan, H.S. Bassiouny, Transitional flow at the venous anastomosis of an arteriovenous graft: potential activation of the ERK1/2 mechanotransduction pathway, J. Biomech. Eng. 125 (2003) 49-61. https://doi.org/10.1115/1.1537737.

[21] B. Ene-Iordache, A. Remuzzi, Blood flow in idealized vascular access for hemodialysis: a review of computational studies, Cardiovasc. Eng. Technol. 8 (2017) 295-312. https://doi.org/10.1007/s13239-017-0318-x.

[22] M. Sarmast, H. Niroomand-Oscuii, F. Ghalichi, E. Samiei, Evaluation of the hemodynamics in straight 6-mm and tapered 6-to 8-mm grafts as upper arm hemodialysis vascular access, Med. Biol. Eng. Comput. 52 (2014) 797-811. https://doi.org/10.1007/s11517-014-1181-7.

[23] G. De Nisco, D. Gallo, K. Siciliano, P. Tasso, M.L. Rizzini, V. Mazzi, K. Calò, M. Antonucci, U. Morbiducci, Hemodialysis arterio-venous graft design reducing the hemodynamic risk of vascular access dysfunction, J. Biomech. 100 (2020) e109591. https://doi.org/10.1016/j.jbiomech.2019.109591.

[24] D. Williams, E.C. Leuthardt, G.M. Genin, M. Zayed, Tailoring of Arteriovenous Graft-to-Vein Anastomosis Angle to Attenuate Pathological Flow Fields, Sci. Rep. 11 (2021) e12153. https://doi.org/10.1038/s41598-021-90813-3.

[25] M.N. Ngoepe, B.D. Reddy, D. Kahn, C. Meyer, P. Zilla, T. Franz, A numerical tool for the coupled mechanical assessment of anastomoses of PTFE arterio-venous access grafts, Cardiovasc. Eng. Technol. 2 (2011) 160-172. https://doi.org/10.1007/s13239-011-0045-7.

[26] S. Mok, S. Cho, J. Lee, J.Y. Kim, S.S. Yun, Y.J. Park, S.C. Park, J. Lee, Optimizing venous anastomosis angle for arteriovenous graft with intimal hyperplasia using computational fluid dynamics, J. Mech. Sci. Technol. 37 (2023) 5231-5238. https://doi.org/10.1007/s12206-023-0925-4.

[27] F. Kabinejadian, B. Su, D.N. Ghista, M. Ismail, S. Kim, H.L. Leo, Sequential venous anastomosis design to enhance patency of arterio-venous grafts for hemodialysis, Comput. Methods Biomech. Biomed. Eng. 20 (2017) 85-93. https://doi.org/10.1080/10255842.2016.1200564.

[28] K. Van Canneyt, U. Morbiducci, S. Eloot, G. De Santis, P. Segers, P. Verdonck, A computational exploration of helical arterio-venous graft designs, J. Biomech. 46 (2013) 345-353. https://doi.org/10.1016/j.jbiomech.2012.10.027.

[29] F. Kabinejadian, M. McElroy, A. Ruiz-Soler, H.L. Leo, M.A. Slevin, L. Badimon, A. Keshmiri, Numerical assessment of novel helical/spiral grafts with improved hemodynamics for distal graft anastomoses, PLoS One 11 (2016) e0165892. https://doi.org/10.1371/journal.pone.0165892.

[30] M.H. Pahlavanian, D.D. Ganji, Computational assessment of venous anastomosis angles and graft configurations in arteriovenous graft, Results Eng. 17 (2023) e100944. https://doi.org/10.1016/j.rineng.2023.100944.

[31] M. Anwer, R.M.C. SO, Swirling turbulent flow through a curved pipe. I: Effect of swirl and bend curvature, Exp. Fluids 14 (1993) 85-96. https://doi.org/10.1007/BF00196992.

[32] S. Quicken, B. Mees, N. Zonnebeld, J. Tordoir, W. Huberts, T. Delhaas, A. Sequeira, A realistic arteriovenous dialysis graft model for hemodynamic simulations, PloS one. 17 (2022) e0269825. https://doi.org/10.1371/journal.pone.0269825.

[33] S. Quicken, Y. de Bruin, B. Mees, J. Tordoir, T. Delhaas, W. Huberts, Computational study on the haemodynamic and mechanical performance of electrospun polyurethane dialysis grafts, Biomech. Model. Mechanobiol. 19 (2020) 713-722. https://doi.org/10.1007/s10237-019-01242-1.

[34] S. Quicken, T. Delhaas, B.M.E. Mees, W. Huberts, Haemodynamic optimisation of a dialysis graft design using a global optimisation approach, Int. J. Numer. Method Biomed. Eng. 37 (2021) e3423. https://doi.org/10.1002/cnm.3423.

[35] L. Grechy, F. Iori, R.W. Corbett, W. Gedroyc, N. Duncan, C.G. Caro, P.E. Vincent, The effect of arterial curvature on blood flow in arterio-venous fistulae: Realistic geometries and pulsatile flow, Cardiovasc. Eng. Technol. 8 (2017) 313-329. https://doi.org/10.1007/s13239-017-0321-2.

[36] B. Berthier, R. Bouzerar, C. Legallais, Blood flow patterns in an anatomically realistic coronary vessel: influence of three different reconstruction methods, J. Biomech. 35 (2002) 1347-1356. https://doi.org/10.1016/S0021-9290(02)00179-3.

[37] S.F.C. Stewart, E.G. Paterson, G.W. Burgreen, P. Hariharan, M. Giarra, V. Reddy, S.W. Day, K.B. Manning, S. Deutsch, M.R. Berman, M.R. Myers, R.A. Malinauskas, Assessment of CFD Performance in Simulations of an Idealized Medical Device: Results of FDA's First Computational Interlaboratory Study, Cardiovasc. Eng. Technol. 3 (2012) 139-160. https://doi.org/10.1007/s13239-012-0087-5.

[38] L. Li, C.M. Terry, Y.-T.E. Shiu, A.K. Cheung, Neointimal hyperplasia associated with synthetic hemodialysis grafts, Kidney Int. 74 (2008) 1247-1261. https://doi.org/10.1038/ki.2008.318.

[39] J.B. Thomas, J.S. Milner, B.K. Rutt, D.A. Steinman, Reproducibility of image-based computational fluid dynamics models of the human carotid bifurcation, Ann. Biomed. Eng. 31 (2003) 132-141. https://doi.org/10.1114/1.1540102.

[40] M. Motomiya, N. Watanabe, M. Ota, K. Shimoda, D. Kawamura, N. Iwasaki, Efficacy of the microscopic parachute end-to-side technique for creating large-to-small venous anastomoses in free flaps in the extremities, J. Plast. Reconstr. Aesthet. Surg. 34 (2022) 189-198. https://doi.org/10.1016/j.jpra.2022.10.003.

[41] U. Krueger, A. Huhle, K. Krys, H. Scholz, Effect of Tapered Grafts on Hemodynamics and Flow Rate in Dialysis Access Grafts, Artif. Organs 28 (2004) 623-628. https://doi.org/10.1111/j.1525-1594.2004.07367.x.

[42] A. Moufarrej, J. Tordoir, B. Mees, Graft modification strategies to improve patency of prosthetic arteriovenous grafts for hemodialysis, J. Vasc. Access 17 (2016) S85-S90. https://doi.org/10.5301/jva.5000526.

[43] H.G. Weller, G. Tabor, H. Jasak, C. Fureby, A tensorial approach to computational continuum mechanics using object-oriented techniques, Comput. Phys. 12 (1998) 620-631. https://doi.org/10.1063/1.168744.

[44] S.A. Wright, F.M. O'Prey, A.L. Bell, G.E. Mcveigh, D.J. Rea, R.D. Plumb, A.J. Gamble, W.J. Leahey, A.B. Devine, C.M. R, D.G. Johnston, M.B. Finch, Microcirculatory hemodynamics and endothelial dysfunction in systemic lupus erythematosus, Arterioscler. Thromb. Vasc. Biol.. 26 (2006) 2281-2287. https://doi.org/10.1161/01.ATV.0000238351.82900.7f.

[45] J.D. Humphrey, M.A. Schwartz, Vascular mechanobiology: homeostasis, adaptation, and disease, Annu. Rev. Biomed. Eng. 23 (2021) 1-27. https://doi.org/10.1146/annurev-bioeng-092419-060810. 10.1146/annurev-bioeng-092419-060810.

[46] S.E. Rittgers, C. Garcia-Valdez, J.T. McCormick, M.P. Posner, Noninvasive blood flow measurement in expanded polytetrafluoroethylene grafts for hemodialysis access, J. Vasc. Surg. 3 (1986) 635-642. https://doi.org/10.1016/0741-5214(86)90289-2.

[47] M. Selmi, H. Belmabrouk, A. Bajahzar, Numerical study of the blood flow in a deformable human aorta, Appl. Sci. 9 (2019) 1216. https://doi.org/10.3390/app9061216.

[48] A.M. Malek, S.L. Alper, S. Izumo, Hemodynamic shear stress and its role in atherosclerosis, J. Am. Med. Assoc. 282 (1999) 2035-2042. https://doi.org/10.1001/jama.282.21.2035.

[49] N. Tran-Nguyen, F. Condemi, A. Yan, S. Fremes, P. Triverio, L. Jimenez-Juan, Wall shear stress differences between arterial and venous coronary artery bypass grafts one month after surgery, Ann. Biomed. Eng. 50 (2022) 1882-1894. https://doi.org/10.1007/s10439-022-03007-x.

[50] I. Van Tricht, D. De Wachter, J. Tordoir, P. Verdonck, Comparison of the hemodynamics in 6 mm and 4–7 mm hemodialysis grafts by means of CFD, J. Biomech. 39 (2006) 226-236. https://doi.org/10.1016/j.jbiomech.2004.12.003.

[51] B. Ene-Iordache, L. Mosconi, G. Remuzzi, A. Remuzzi, Computational fluid dynamics of a vascular access case for hemodialysis, J. Biomech. Eng. 123 (2001) 284-292. https://doi.org/10.1115/1.1372702.

[52] F. Yilmaz, M.Y. Gundogdu, A critical review on blood flow in large arteries; relevance to blood rheology, viscosity models, and physiologic conditions, Korea-Australia Rheology Journal 20 (2008) 197-211.

[53] J. Chen, X.-Y. Lu, W. Wang, Non-Newtonian effects of blood flow on hemodynamics in distal vascular graft anastomoses, Journal of Biomechanics 39 (2006) 1983-1995.

[54] A. Quarteroni, A. Manzoni, C. Vergara, The cardiovascular system: mathematical modelling, numerical algorithms and clinical applications, Acta Numerica 26 (2017) 365-590. https://doi.org/10.1017/S0962492917000046.